\documentclass[12pt,preprint,psfig]{aastex}

\def\ltsima{$\; \buildrel < \over \sim \;$}

\def\lsim{\lower.5ex\hbox{\ltsima}}
\def\gtsima{$\; \buildrel > \over \sim \;$}
\def\gsim{\lower.5ex\hbox{\gtsima}}

\begin{document}

\title{Dwarf Galaxy Formation Was Suppressed By Cosmic Reionization}

\author{J. Stuart B. Wyithe\altaffilmark{1} and Abraham
Loeb\altaffilmark{2}}


\altaffiltext{1}{University of Melbourne, Parkville, Victoria, Australia}

\altaffiltext{2}{Harvard-Smithsonian Center for Astrophysics, 60 Garden
St., Cambridge, MA 02138}

{\bf A large number of faint galaxies, born less than a billion years
after the big bang, have recently been discovered~$^{1-6}$.  The
fluctuations in the distribution of these galaxies contributed to a
scatter in the ionization fraction of cosmic hydrogen on scales of
tens of Mpc, as observed along the lines of sight to the earliest
known quasars~$^{7-9}$.  Theoretical simulations predict that
the formation of dwarf galaxies should have been suppressed after
cosmic hydrogen was reionized~$^{10-13}$, leading to a
drop in the cosmic star formation rate~$^{14}$.  Here we present
evidence for this suppression. We show that the post-reionization
galaxies which produced most of the ionizing radiation at a redshift
$z\sim 5.5$, must have had a mass in excess of
$\sim10^{10.6\pm0.4}M_\odot$ or else the aforementioned scatter would
have been smaller than observed.  This limiting mass is two orders of
magnitude larger than the galaxy mass that is thought to have
dominated the reionization of cosmic hydrogen ($\sim10^8M_\odot$). We
predict that future surveys with space-based infrared telescopes will
detect a population of smaller galaxies that reionized the Universe at
an earlier time, prior to the epoch of dwarf galaxy suppression.}

The lines of sight towards the earliest known quasars at redshifts $z>6$
show a rapid transition in the ionization state of the intergalactic medium
(IGM), potentially marking the end of the reionization
epoch~$^{15-18}$.  It is
thought that the dominant source population to have triggered reionization
included dwarf galaxies with virial temperatures above the hydrogen cooling
threshold ($T_{\rm vir}\sim10^4$K). The re-ionization of cosmic hydrogen
resulted in heating of the IGM to $\sim10^4$K which drastically increased
the minimum virial temperature of new galaxies~$^{12-13}$ to
a value as high as $T_{\rm vir}\sim10^5$K. Reionization is therefore
expected to be followed by a depletion of low-mass galaxies and hence a
decrease in the global star formation rate~$^{14}$. The detection of
this anticipated suppression at $z\sim 6$ would provide important evidence
for a late reionization epoch.

In addition to luminous quasars, starburst galaxies with star formation
rates in excess of $\sim0.1 M_\odot~{\rm yr}^{-1}$ and dark matter
halos~$^{6}$ of $\sim 10^{9-10}M_\odot$ have recently been
discovered~$^{1-6,19-22}$
at $z\sim5$--$6$. These sources contributed to the ionizing background
radiation at the end of the reionization epoch. Luminous Ly$\alpha$
emitters are routinely identified through continuum dropout and narrow band
imaging techniques~$^{3,4,20}$. However, in
order to study fainter sources which were potentially responsible for
reionization, spectroscopic searches have been undertaken near the critical
curves of lensing galaxy
clusters~$^{21-23}$, where gravitational
magnification enhances the flux sensitivity.  A comparison between the
findings of these deep surveys and the statistics of bright Ly$\alpha$
emitters~$^{22}$ provides a preliminary hint for the depletion of low
mass galaxies, as expected from galaxy formation in a photo-ionized
IGM~$^{12-13}$. However, this inference relies on a
comparison between heterogeneous data sets based on different methods of
discovery.

The upper right panel of Figure~\ref{fig1} shows data on the luminosity
function of galaxies at $z\sim$5.5-6 from the Hubble Space Telescope
Ultra-Deep Field (UDF)~$^{4}$.  We first use this data to constrain
the physical characteristics of high redshift galaxies (see \S~1 of the
{\em Supplementary Online Materials}). We model starburst galaxies based on
a simple prescription for their continuum luminosity~$^{24}$ combined
with the Sheth-Tormen~$^{25}$ mass-function of galaxy halos. The two
free parameters in this model are the starburst lifetime ($t_{\rm lt}$) and
the mass fraction of virialized baryons which are converted into stars in
each galaxy ($f_{\rm star}$). The lower right panel shows contours of the
joint a-posteriori probability distribution $[{d^2P}/{d(\log{f_{\rm star}})d(\log{t_{\rm
lt}})}]$ (see \S~2 of the {\em Supplementary Online Materials} for a
description).  The best-fit model luminosity function is plotted over the
data, with the favored values of $t_{\rm lt}$ and $f_{\rm star}$
listed. The other contours (grey) show the virial temperature corresponding
to the lowest luminosity bin [labeled by log$_{10}(T_{\rm vir})$]. The
best-fit luminosity function implies that the faintest observed galaxies
have a virial temperature $\sim10^{5}$K, and that starburst galaxies at
$z\sim6$ have duty-cycles of $t_{\rm lt}/t_{\rm H}\sim10\%$ (where
$t_{\rm H}$ is the Hubble time), combined with star formation efficiencies
of $f_{\rm star}\sim10-15\%$. Interestingly, a duty-cycle of $t_{\rm
lt}/t_{\rm H}\sim0.1$ corresponds to the dynamical time at the virial
overdensity of 200, i.e. the time it takes the virialized gas to settle to
the center of a collapsing galaxy. For the purpose of understanding the
process of galaxy formation at high redshifts and the way it shapes
reionization, it is crucial to know whether star formation was suppressed
in dwarf galaxies near the hydrogen cooling threshold. With a virial
temperature of $T_{\rm vir}\sim10^4$K, these galaxies had a total (halo)
mass of $\sim10^8M_\odot$ at $z\sim 6$. Unfortunately, present detection
limits are insufficient to probe the existence of such galaxies.

We overcome the limitation of direct detection of faint galaxies by
making use of the fact that the absorption spectra of high redshift
quasars reveal large fluctuations in the IGM ionization state on very
large scales~$^{7-9,17}$
(tens to hundreds of Mpc). While observations of the galaxy luminosity
function are flux limited, the fluctuations in the ionizing background
are sensitive to contributions from galaxies of all luminosities. A
measurement of the fluctuations in the ionizing background therefore
provides a powerful probe of the properties of the galaxies generating
the ionizing background radiation at the end of the reionization era,
even if those galaxies themselves lie below current detection
limits. The background fluctuations can be inferred from observed
fluctuations in the effective Ly$\alpha$ optical depth, $\tau_{\rm
eff}$. Fan et al.~$^{9}$ have measured average values for the
effective optical depths of $\tau_{\rm eff}=2.5$, $2.6$, $3.2$ and
$4.0$, within redshift bins of width $\Delta z=0.15$ at $z=5.25$,
$5.45$ 5.65 and 5.85, respectively. At these redshifts the scatter in
$\tau_{\rm eff}$ among the different lines-of-sight is
$\sigma_\tau=0.5$, $0.6$, $0.8$ and $0.8$, respectively. The dominant
uncertainty in individual estimates of $\tau_{\rm eff}$ is the level
of unabsorbed quasar continuum, which introduces noise~$^{9}$ in
$\tau_{\rm eff}$ of $\pm0.05$. This observational uncertainty must be
subtracted (in quadrature) from $\sigma_\tau$ to obtain an estimate of
the intrinsic scatter in $\tau_{\rm eff}$. Following this operation we
find fractional fluctuations $\delta_\tau\equiv\sigma_\tau/\tau_{\rm
eff}$ of $(0.20\pm0.03)$, $(0.23\pm0.03)$, $(0.25\pm0.03)$, and
$(0.20\pm0.03)$ at the above sequence of redshifts. The corresponding
comoving mean-free-paths of ionizing photons are $R_{\rm
mfp}=(39^{+16}_{-10})$Mpc, $(41^{+21}_{-15})$Mpc,
$(27^{+21}_{-10})$Mpc and $(19^{+10}_{-15})$Mpc
respectively~$^{9}$. Here we have corrected the frequency averaged
ionization cross-section for a spectrum corresponding to a
star-forming galaxy~$^{26}$. At the highest redshift under
consideration, some lines-of-sight contain blank absorption troughs
making their estimated value of $\tau_{\rm eff}$ a lower limit. At
$z\sim5.85$ there is no measured lower limit for $R_{\rm mfp}$, and so
we conservatively set it to 3Mpc (below the typical value at
$z\sim6$).

We have calculated the predicted scatter in $\tau_{\rm eff}$ from a
combination of cosmic variance among different regions and Poisson noise in
the number of ionizing sources, based on an extension of a detailed model
published elsewhere~$^{27}$. A description of this model is provided in
\S~3 of the {\it Supplementary Online Materials}, and only a summary of its
main features is given here.  Fluctuations in both the ionizing background
intensity, $J$, and the density field around their mean values, contribute
to fluctuations in $\tau_{\rm eff}$.  Within our model, the ionizing
background is smoothed on the scale $R_{\rm mfp}$, since any given point in
the IGM is exposed to all sources within a volume $V_{\rm mfp}=(4\pi/3)
R_{\rm mfp}^3$ around it. The model combines four major sources of
fluctuations: $(i)$ Poisson noise in the number of sources contained within
a volume $V_{\rm mfp}$; $(ii)$ Fluctuations in the value of $J$ within
volumes of $V_{\rm mfp}$ resulting from delayed (or enhanced) structure
formation in underdense (overdense) regions~$^{28}$. In evaluating this
quantity we include two contributions to the derivative $dJ/dz$ arising
from: {\it (a)} the evolution of the star formation rate, which at first we
take to be proportional to the derivative of the mass fraction of baryons
that were assembled into galaxies; and {\it (b)} the evolution of $R_{\rm
mfp}$; $(iii)$ Fluctuations in Ly$\alpha$ transmission due to fluctuations
in the density contrast smoothed on the scale $R_{\rm mfp}$; $(iv)$
Fluctuations in the transmission due to small-scale density fluctuations
along the line-of-sight through a region with constant $J$. These latter
fluctuations are computed based on the results of numerical
simulations~$^{29}$.

We have compared the predictions of the above model with observational
data as a function of the lifetime $t_{\rm lt}$ and the minimum virial
temperature $T_{\rm min}$ of the lowest mass galaxies that make a
major contribution to the ionizing background.  Contours of the
resulting distribution ${d^2P}/{d(\log{t_{\rm lt}})d(\log{T_{\rm
min}})}$ (see \S~4 of the {\it Supplementary Online Materials} for
details) evaluated at $z\sim5.45$ (blue contours) and $z\sim5.65$ (red
contours) are plotted in the lower left panel of
Figure~\ref{fig1}. The corresponding marginalized differential
probability distributions $dP/d(\log{T_{\rm min}})$ are plotted in the
upper left panel (solid blue and red lines). In addition we plot
$dP/d(\log{T_{\rm min}})$ at $z\sim5.25$ (dashed blue line) and
$z\sim5.85$ (dashed red line). Intriguingly, the results at $z\la5.65$
favor $T_{\rm min}\ga10^5$K, while results at $z\sim5.85$ favor
$T_{\rm min}\sim 10^{4-5}$K [though note that $\sigma_\tau(z=5.85)$ is
a lower limit].  This trend is expected if reionization completed at
$z\sim6$--7. Indeed, the mini-halos ($T_{\rm vir}<10^4$K) which
provide the collapsed gas reservoir for star formation in galaxies
above the hydrogen cooling threshold are thought to be
photo-evaporated~$^{30}$ on a timescale of $\sim0.2t_{\rm H}$,
which corresponds to $\Delta z\sim 1$. Thus, if reionization completed
at $z\sim6.5-7$, we would only expect to observe the strong effect of
dwarf galaxy suppression at $z\la5.5-6$. We also plot the combined
differential and cumulative distributions using data from the three
redshift bins below $z\sim5.75$ (black lines), and find values of
$T_{\rm min}\sim10^{5.5\pm0.3}$K (68\%), with $T_{\rm min}\la10^{4}$K
ruled out at a (statistical) confidence level of $>99\%$.  The above
constraints have conservatively assumed a prior probability
distribution for $T_{\rm min}$ that extends down to $T_{\rm
min}=10^3$K. On the other hand, the ratio between the probabilities of
having $10^4\mbox{K}<T_{\rm min}<10^{5}\mbox{K}$ and
$10^{5}\mbox{K}<T_{\rm min}<10^{6}\mbox{K}$ is independent of any
assumed lower cutoff. We find this ratio to be $\sim0.05$. A limiting
virial temperature of $T_{\rm min}\ga10^5$K is an order of magnitude
above the threshold for star formation triggered by atomic hydrogen
cooling. However, $T_{\rm min}\ga10^5$K is consistent with galaxy
formation in an IGM that was heated by UV photo-ionization at a higher
redshift, thus preventing starbursts in low-mass galaxies from
dominating the global star formation rate by $z\sim5.5$.

Our analysis so far has assumed for simplicity that the star formation
rate is proportional to the derivative of the collapsed fraction in
halos with $T_{\rm vir}>T_{\rm min}$. However, starburst episodes are
believed to be triggered by galaxy mergers. We have therefore repeated
the above calculation, replacing the derivative of the collapsed
fraction with a star formation model~$^{14}$ that combines the
contribution from mergers of halos with an initial $T_{\rm vir}<T_{\rm
min}$ (that go above $T_{\rm min}$ after the merger) and the
contribution from mergers of galaxies with an initial $T_{\rm
vir}>T_{\rm min}$ (see \S~5 of the {\it Supplementary Online
Materials} for details). The results are shown in Figure~\ref{fig1s} of the {\em
Supplementary Online Materials}. This alternative estimate of the star
formation rate also leads to a preferred value of $T_{\rm
min}\sim10^{5.5\pm0.3}$K, with $T_{\rm min}\ga10^{4}$K at the
$\ga99\%$ confidence level.

In the local Universe, star formation within galaxies with stellar masses
below $\sim10^{10}M_\odot$ is suppressed as a result of supernova-driven
winds which expel gas from their shallow potential
wells~$^{31-32}$. At high redshifts the lifetime of massive
stars could be comparable to the halo dynamical time and this feedback may
not operate as effectively. Nevertheless, it is important explore the
effect of such a feedback on fluctuations in the ionizing background and
hence on our conclusions regarding the suppression of dwarf galaxy
formation in the reionized IGM.  We have repeated our analysis of the
galaxy luminosity function, with an allowance for a suppression of star
formation in galaxies below a critical mass (which is treated as a free
parameter).  Details of the required modifications and the results are
described in \S~6 and Figure~\ref{fig2s} of the {\it Supplementary Online
Materials}. Allowing for feedback in high redshift galaxies leads to
estimates of $f_{\rm star}\sim0.5$, with a duty-cycle that is larger
than $t_{\rm lt}/t_{\rm H}\sim0.5$. The best fit value for the critical
mass scale below which supernovae feedback operates is
$\sim2\times10^{10}M_\odot$, or $T_{\rm vir}\sim10^{5}K$. 

Estimates of the scatter in the ionizing background due to Poisson noise in
the number of galaxies and due to fluctuations in the star formation rate
as a result of delayed or enhanced structure formation must also be
modified to account for possible supernova feedback. We find that if
supernova feedback operates at high redshift in the same way as observed at
low redshift, then the most likely value is $T_{\rm min}\sim10^{5.5^{+0.3}_{-0.4}}$K in
agreement with our previous calculation, while $T_{\rm min}\la10^4$K is ruled
out at the $\ga95\%$ level. Note that the critical mass-scales derived for
supernovae feedback and $M_{\rm min}$ are similar, although they were
derived from different data sets and astrophysical considerations. However,
our results imply that a sharp cutoff due to reionization is required to
explain the observed scatter in the effective optical depth of the IGM in
addition to any suppression resulting from supernovae feedback.

The observed star formation rate in the post-reionization era is on the
borderline of being sufficient to reionize the Universe~$^{4}$.  The
value of $T_{\rm min}\ga10^5$K at $z\sim5.5$ happens to describe the
faintest galaxies in the UDF. We therefore predict that star formation is
suppressed in galaxies just below the current limit of galaxy surveys at
$z\sim6$.  Future searches at higher redshifts with JWST or large-aperture
ground-based infrared telescopes should find a population of galaxies with
$T_{\rm vir}\sim 10^4$K and detect an enhanced star formation rate prior to
completion of reionization~$^{14}$.

\vskip 0.2in

\bigskip
\bigskip
\noindent
{\bf Acknowledgments:}\\ The authors wish to thank Adam Lidz for
helpful discussions regarding this work. Our research was supported in
part by grants from ARC, NSF and NASA.

\bigskip
\bigskip
\noindent
{\bf Author Information:}\\ Correspondence and requests for materials to Stuart Wyithe (swyithe@physics.unimelb.edu.au) or Abraham Loeb (aloeb@cfa.harvard.edu).

\newpage

\begin{figure*}[t]
\epsscale{1.}  \plotone{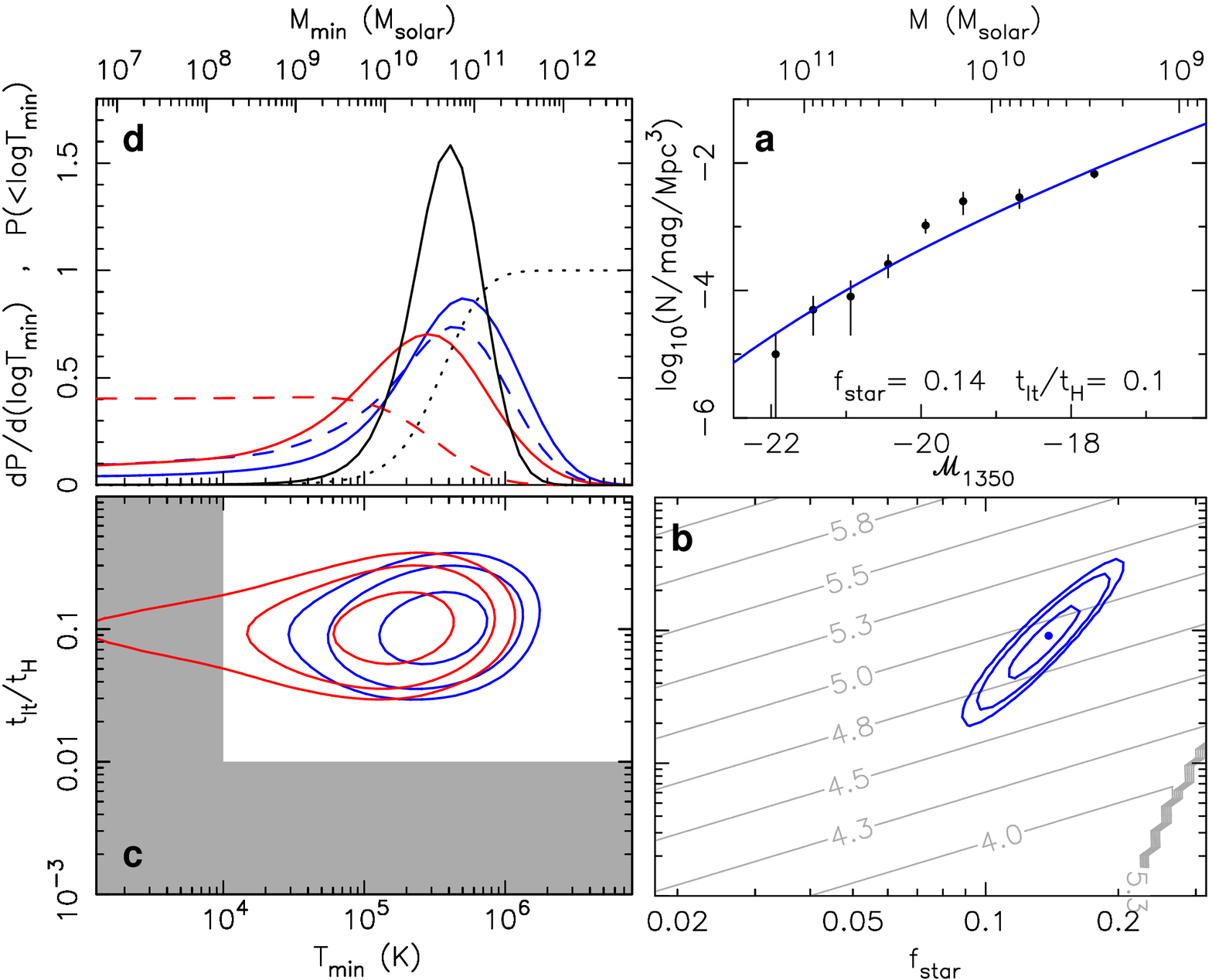}
\end{figure*}
\begin{figure*}[t]
\caption{\label{fig1} Constraints on the star-formation efficiency ($f_{\rm
star}$), duty-cycle ($t_{\rm lt}/t_{\rm H}$) and minimum virial temperature
($T_{\rm min}$) of galaxies at $z\sim5.5-6$. {\bf Panel~a:} The
luminosity function inferred from the UDF~$^{4}$, together with the
best-fit solution for a luminosity function model based on the Sheth-Tormen
~$^{25}$ mass function (with luminosity proportional to halo mass).
$\mathcal{M}_{1350}$ is the galaxy's absolute AB magnitude at a rest-frame
wavelength of 1350\AA. The upper axis shows the halo mass corresponding to
$\mathcal{M}_{1350}$ for the best fit.
{\bf Panel~b:} Contours (at 64, 26, and 14\% of the maximum
likelihood) of the joint a-posteriori probability distribution
$[{d^2P}/{d(\log{f_{\rm star}})d(\log{t_{\rm lt}})}]$. For comparison, the grey contours
show the virial temperature of halos [labeled using $\log_{10}(T_{\rm
vir})$] corresponding the lowest luminosity bin in the UDF. {\bf 
Panel~c:} The joint probability distributions for $\log(t_{\rm lt}/t_{\rm H})$ and
$\log(T_{\rm min})$, which include the constraint on $t_{\rm lt}/t_{\rm H}$ from
the galaxy luminosity function. The contours represent values at 64, 26 and
14\% of the maximum likelihood, and are shown for observations at
$z\sim5.45$ (blue contours) and $z\sim5.65$ (red contours). This
distribution includes constraints on $t_{\rm lt}$ from the galaxy
luminosity function. However assuming the most likely value of $R_{\rm
mfp}$, we find that $T_{\rm min}\ga10^{5}$K at $z\la5.5$ in all cases where
the duty-cycle of the UV-bright phase of galaxies is greater than 1\%
(corresponding to the lifetime of massive stars). {\bf Panel~d:}
Differential probability distributions for $\log{(T_{\rm min})}$ at $z\sim5.25$
(dashed blue lines), $z\sim5.45$ (blue lines), $z\sim5.65$ (red lines) and
$z\sim5.85$ (dashed red lines). We also show combined constraints using the
lowest three redshift bins (black lines). For comparison, the upper axis
shows the corresponding masses at $z=5.5$. In this case the solid and
dotted lines correspond to differential and cumulative distributions.
Throughout this {\it Letter}, we adopt the latest values for the
cosmological parameters as inferred from the {\em Wilkinson Microwave
Anisotropy Probe} data~$^{33}$.}
\end{figure*}

\clearpage

\title{\bf SUPPLEMENTARY ONLINE MATERIAL}

\noindent The above {\em Letter to Nature} 
interprets the observed luminosity function of $z\sim 6$ galaxies and the
large-scale fluctuations in the Ly$\alpha$ optical depth towards $z\sim 6$
quasars, based on a detailed theoretical model. Below we provide full
details on the basic version of this model as well as its variants. The
following sections expand the brief descriptions that can be found in the
original {\em Letter to Nature}.

\section{Modeling the galaxy luminosity function}
\label{LF}

The properties of the observed starburst galaxies are interpreted through a
comparison with a model luminosity function.  We employ two estimates of the
relationship between the luminosity of a starburst and its host halo
mass. Our fiducial model assumes that the starburst luminosity $L$ is
proportional to halo mass $M$, i.e. $L\propto M$. In this case, the
luminosity function $\Psi$ (with units of number of galaxies per magnitude
per comoving volume) may be written as
\begin{equation}
\label{LF}
\Psi = \frac{t_{\rm lt}}{t_{\rm H}}\frac{dn}{dM}\frac{dM}{d\mathcal{M}_{1350}},
\end{equation}
where $\mathcal{M}_{1350}(L)$ is the starburst luminosity expressed in AB
magnitudes at a rest-frame wavelength of 1350 \AA.  Following Loeb et
al.~(2005), we express the relation between luminosity and halo mass
through two free parameters: {\it (i)} the mass fraction of virialized
baryons converted into stars (the so-called ``star formation efficiency''),
$f_{\rm star}$; and {\it (ii)} the fraction of the Hubble time during which
the starburst emits most of its ionizing radiation, $t_{\rm lt}/t_{\rm H}$
(the so-called ``duty-cycle'').

\section{A-posteriori constraints from the luminosity function}

Using the models described above, the joint a-posteriori probability
distribution $\frac{d^2P}{df_{\rm star}dt_{\rm lt}}$ for $f_{\rm star}$ and
$t_{\rm lt}$ was calculated as $\frac{d^2P}{df_{\rm star}dt_{\rm lt}}
\propto L_{f_{\rm star},t_{\rm lt}} \frac{dP_{\rm prior}}{df_{\rm
star}}\frac{dP_{\rm prior}}{dt_{\rm lt}}$, where $L_{f_{\rm
star},t_{\rm lt}}$ is the likelihood for the combination of $f_{\rm
star}$ and $t_{\rm lt}$ given the observed luminosity function.  Here
we have taken the prior probability distributions for $f_{\rm star}$
and $t_{\rm lt}$ to be flat in the logarithm, i.e. $dP_{\rm
prior}/dx\propto 1/x$. The best fit luminosity function, and
constraints on $f_{\rm star}$ and $t_{\rm lt}/t_{\rm H}$ are shown in
the right hand panels of Figure~\ref{fig1s} of this supplement for the
fiducial model (repeating results presented in Figure 1 of the
original {\em Letter to Nature}). Marginalizing over $f_{\rm star}$ in
this distribution provides a probability distribution for $t_{\rm
lt}/t_{\rm H}$ which is then used in the evaluation of limits on the
minimum virial temperature $T_{\rm vir}$ of high-redshift galaxies.

\section{Scatter in the effective optical depth along lines of sight
towards different quasars at $z\ga 6$}

At a fixed redshift $z$, the local value of Ly$\alpha$ optical depth
($\tau_{\rm l}$) scales as
\begin{equation}
\tau_{\rm l} \propto \rho_{\rm HI}^2 J^{-1},
\end{equation}
where $\rho_{\rm HI}$ is the density of neutral hydrogen and $J$ is
the ionizing background intensity. However the effective optical depth
($\tau_{\rm eff}$) is measured across some path-length through the
intergalactic medium (IGM) and must be computed from $\tau_{\rm
eff}=-\log{F}$ where $F=\langle e^{-\tau_{\rm l}}\rangle$ is the
transmission averaged over the path-length. While the ionizing
background will be sensitive to cosmic variance in density contrast
calculated in 3-dimensional volumes ($\Delta$) on the scale of the
mean-free-path for ionizing photons $R_{\rm mfp}$, the absorption
spectra of quasars probe only a narrow cylinder (skewer) through these
volumes. Note that although $\tau_{\rm l}$ varies with $z$ due to the
expansion of the universe and the growth factor of density
perturbations, we focus here on the scatter in $\tau_{\rm eff}$ at a
fixed $z$.

We calculate the predicted scatter in $\tau_{\rm eff}$ using a combination
of cosmic variance among different regions and Poisson noise in the number
of ionizing sources. Our calculation is based on an extension of a detailed
model that we published in Wyithe \& Loeb~(2006). A full description of the
methodology of the model is given in that paper. Here we give a summary,
and point out the various additions and improvements. Fluctuations in both
the ionizing background intensity $J$ about the mean value $\bar{J}$ as
well as inhomogeneities in the gas density, contribute to fluctuations in
$\tau_{\rm eff}$.  The ionizing background is smoothed on the scale $R_{\rm
mfp}$, since any point in the IGM sees sources within a volume $V_{\rm
mfp}=(4\pi/3) R_{\rm mfp}^3$. The model combines four major sources of
fluctuations in $\tau_{\rm eff}$:

\vspace{5mm}
\noindent $(i)$ {\em Poisson noise in the number of sources
contained within a volume $V_{\rm mfp}$} 

The scatter in the radiation field due to the finite numbers of
galaxies is computed by weighting the variance in the number of
sources by luminosity, i.e.
\begin{equation}
\sigma_{\rm J} = \frac{\sqrt{\int_{\mathcal{M}_{\rm 1350,max}}^{-\infty} d\mathcal{M}_{1350} \left[L(\mathcal{M}_{1350}) \sqrt{V_{\rm mfp}\Psi(\mathcal{M}_{1350} )}\right]^2}  }  {\int_{\mathcal{M}_{\rm 1350,max}}^{-\infty} d\mathcal{M}_{1350} L(\mathcal{M}_{1350}) V_{\rm mfp}\Psi(\mathcal{M}_{1350}) },
\end{equation}
where $L$ is the luminosity, and
$\mathcal{M}_{\rm 1350,max}$ is the absolute AB magnitude at rest-frame
1350\AA  corresponding to the minimum virial temperature $T_{\rm min}$.  We
define fluctuations in $J$ due to Poisson noise to be $\delta_{J,{\rm
P}}\equiv (J-\bar{J})/\bar{J}$;

\noindent $(ii)$ {\em Fluctuations in the value of
$J$ within volumes of $V_{\rm mfp}$ resulting from delayed or enhanced
structure formation in underdense or overdense regions} 

Large-scale inhomogeneity in the cosmic density field leads to
structure-formation that is enhanced in over-dense regions and delayed
in under-dense regions.  The resulting cosmic variance in the redshift
at which a critical value of collapsed fraction ($F_{\rm col}$) is
reached within regions of size $R_{\rm mfp}$ may be calculated as
\begin{equation}
\label{delta}
\langle\delta z_{\rm cv}^2\rangle^{\frac{1}{2}} =
\frac{\sigma(R_{\rm mfp})}{\delta_c}(1+z),
\end{equation}
where $\sigma(R_{\rm mfp})$ is the r.m.s. amplitude of the linear density
field smoothed over spheres of radius $R_{\rm mfp}$, and $\delta_c$ is a
critical overdensity for collapse [$\propto (1+z)$ at high redshift]. Since
the ionizing background within a given region depends on the value of the
collapsed fraction, we can relate $\langle\delta z_{\rm
cv}^2\rangle^{\frac{1}{2}}$ to the cosmic variance in redshift where a
critical value of $J$ is obtained.  We would like to convert this scatter
in redshift to a scatter in the optical-depth to Ly$\alpha$ absorption at a
given redshift.  In a smooth (i.e. non-clumpy) IGM the fractional change in
the value of the optical depth relative to the average, given a density
contrast ($\Delta_z$), and a delay ($\delta z$) in the redshift where the
critical collapse fraction is reached is
\begin{equation}
\delta_\tau\equiv\frac{\tau-\tau_{\rm av}}{\tau_{\rm av}} = \Delta_z^2 \frac{J(z)}{J(z)+\frac{dJ}{dz}\delta z}-1= \Delta_z^2 \left(1+\frac{d\ln J}{dz}\delta z\right)^{-1}-1
\end{equation}
where $\tau_{\rm av}$ is the average value of optical depth at $z$. Within
our fiducial model the production rate of ionizing photons within a
volume $V_{\rm mfp}$, and hence $J$, is taken to be proportional to
the time-derivative of the fraction of baryons which collapsed into
galaxy halos ($F_{\rm col}$). There are two contributions to the
derivative $dJ/dz$: {\it (a)} the evolution of the star formation
rate; and {\it (b)} the evolution of the co-moving mean-free path. Fan
et al.~(2006) present measurements of $R_{\rm mfp}$ at redshifts
between 4.5 and 5.65, from which we find $R_{\rm
mfp}\propto(1+z)^\gamma$ with $\gamma=-1.75\pm0.75$. Hence in our
fiducial model we have
\begin{equation}
\frac{d\ln J}{dz}=\left(\frac{d\ln(dF_{\rm col}/dt)}{dz}-\frac{\gamma}{(1+z)}\right)\delta z,
\end{equation}
and fluctuations in $J$ due to cosmic variance of
\begin{equation}
\delta_{J,{\rm cv}}\equiv {J-\bar{J}\over \bar{J}}=
\left(\frac{d\ln(dF_{\rm col}/dt)}{dz}-\frac{\gamma}{(1+z)}\right)\delta z.
\end{equation}

\noindent $(iii)$ {\em Fluctuations in Ly$\alpha$ transmission due to
fluctuations in the density contrast smoothed on the scale $R_{\rm
mfp}$}. 

To estimate the typical fluctuations in the effective optical depth in a
smooth IGM we must compute the variance in overdensity among lines of sight
through the density field of length $R_{\rm mfp}$ in addition to
fluctuations in $J$. We calculate the power-spectrum of fluctuations in
cylinders of length $L=2R_{\rm mfp}$ and radius $R$ 
\begin{equation}
P_{\rm 1d}(k) = \frac{1}{(2\pi)^2}\frac{2\pi}{L}\int_k^\infty dy yP_{\rm b}(y)e^{-(y^2-k^2)R^2/4}, 
\end{equation}
where $P_{\rm b}(k)$ as a function of wavenumber $k$ is the linear baryonic
power-spectrum, which may be approximated as $P_{\rm b}(k)=P(k)(1+k^2R_{\rm
f}^2)^{-2}$ in terms of the cold dark matter power-spectrum $P(k)$ and the
filtering scale $R_{\rm f}$ for the associated reionization history. The
variance in $\Delta_z$ on a scale $L$ with wavenumber $k_{\rm L}$ follows
from
\begin{equation}
\sigma_{L,z} = \frac{1}{2\pi}\int_0^{k_{\rm L}}dkP_{\rm 1d}(k).
\end{equation}
The radius of the cylinder $R$ is set by the size of the quasar
emission region, which is much smaller than the filtering scale.  The
fluctuations in the line-of-sight density field are $\delta_{\rm
d}\equiv (\Delta_z-1) = D\delta_z$, where $D$ is the growth factor at $z$ and
$\delta_z$ (with variance $\sigma_{L,z}$) is the overdensity smoothed in thin cylinders of
length $2R_{\rm mfp}$;

\noindent $(iv)$ {\em Fluctuations in the transmission due to small-scale density
fluctuations along the line-of-sight through a region with constant
$J$}

In addition to the smoothed density field averaged on a large scale,
fluctuations in the transmission are sensitive to small scale
structure along the line of sight. Inclusion of these non-linear
effects requires a numerical simulation.  The fluctuations in
effective optical depth $\tau_\Delta$ about a mean $\bar{\tau_\Delta}$
have been computed by Lidz et al.~(2006) based on the
transmission power-spectrum, giving
$\tau_\Delta=2.5^{+0.46}_{-0.32},2.6^{+0.48}_{-0.33},
3.2^{+0.67}_{-0.39}$ and $4.0^{+0.99}_{-0.48}$ respectively, in
regions of co-moving length $L=55$Mpc. We extrapolate from these
results to compute the fluctuations in transmission through thin
cylinders of length $2R_{\rm mfp}$. The transmission ($F$) is related
to the effective optical depth through $F=e^{-\tau_{\rm eff}}$. Hence,
the lower and upper fluctuations in transmission ($\sigma_F$) are
related to the upper and lower fluctuations in optical depth
($\sigma_\tau=\tau_\Delta\delta_\tau$) through the relations
$\sigma_F=e^{-\tau_{\rm eff}}(1-e^{-\sigma_\tau})$ and
$\sigma_F=e^{-\tau_{\rm eff}}(e^{\sigma_\tau}-1)$.  On scales larger
than $\sim10$Mpc the transmission power-spectrum is close to white
noise, and $\sigma_F/F\propto L^{-1/2}$. To estimate
fluctuations in $\tau_\Delta$ within thin cylindrical regions of
length $2R_{\rm mfp}$, we therefore multiply the simulated
transmission fluctuations ($\sigma_F/F$) by $\sqrt{L/(2R_{\rm mfp})}$
(treating upper and lower limits separately).

We assume that each line-of-sight samples a probability distribution
$P(\Delta)$ of density contrasts $\Delta$ relative to the mean density
(as described by Miralda-Escude et al.~2000, ApJ, 530, 1) on scales
$\ll R_{\rm mfp}$.  Defining the local optical depth to be
$\tau_l\equiv A\Delta^2$, one can find an expression
$\tau_{\Delta}=\tau_{\Delta}(A)$ for $\tau_{\Delta}$ given $A$ and
$P(\Delta)$ (Songaila \& Cowie, 2002, Astron. J., 123, 2183;
Lidz et al.~2006).  On the scale $R_{\rm mfp}$, we modulate $A$ based
on the fluctuations in density and radiation,
$A^\prime=A(1+\delta_{\rm d})^2/(1+\delta_{J,{\rm P}} + \delta_{J,{\rm
cv}})$, and hence find the modified
$\tau_{\Delta}^\prime=\tau_{\Delta}^\prime(A^\prime)$. A distribution
of $\tau_{\rm eff}$ that includes fluctuations in $J$ is then found
via Monte-Carlo realizations of the values for $\tau_{\Delta}$,
$\delta_{J,{\rm P}}$, $\delta_{J,{\rm cv}}$ and $\delta_{\rm d}$.

Points in the IGM which are separated by $2R_{\rm mfp}$ have ionizing
backgrounds that are dominated by different sets of sources.  Since the
two-point correlation function is much smaller than unity on the scale
$2R_{\rm mfp}$ (tens of Mpc), we consider regions of this scale to be
independent and assume that the transmission power-spectrum remains close
to white noise in the presence of a radiation field that fluctuates on
scales larger than $2R_{\rm mfp}$.  The number of independent regions in
the ionizing background within a redshift interval $\delta z$ is $N_{\rm
mfp}=\delta z ({cdt}/{dz})/(2R_{\rm mfp})$, where $c$ is the speed of
light. We find $N_{\rm mfp}\sim (1.45_{-0.40}^{+0.55})$,
$(1.35_{-0.45}^{+0.70})$, $(1.9_{-0.8}^{+1.0})$ and $(2.7_{-1.0}^{+9.0})$
at $z=5.25$, $5.45$, $5.65$ and $5.85$ respectively.  To relate modeled
fluctuations in regions of radius $R_{\rm mfp}$ to observations made in
redshift bins, we divide the modeled fluctuations in transmission through
regions of volume $V_{\rm mfp}$ by $\sqrt{N_{\rm mfp}}$.  Finally, to
facilitate comparison of modeled distributions with separate upper and
lower fluctuations (corresponding to an asymmetric distribution) and the
quoted variance in the observed values, we take the root-mean-square of the
upper and lower fluctuations in $\delta_\tau$.

Note that our constraints assume that galaxies dominate $J$. On the other
hand, quasars are more luminous and rare in comparison with galaxies and
may therefore contribute to the fluctuations in the ionizing
background. However, the known population of quasars cannot contribute to
the scatter among regions of size $V_{\rm mfp}$, since there is only $\sim
1$ quasar as bright as the SDSS quasars [or the faintest known X-ray quasar
(Barger et al.~ 2003, ApJ, 584, L61)] at $z\sim6$ per $10^5$ (or $100$)
regions of radius $R_{\rm mfp}\sim20\mbox{Mpc}$. Moreover, the contribution
from quasars to reionization is thought to be much smaller than is required
to reionize the universe at $z\sim5.5$--$6$. Even if there was one dominant
quasar per $V_{\rm mfp}$, it would need to contribute $\ga10\%$ of the
ionizing background in order to influence our results.

\section{A-posteriori constraints on $T_{\rm min}$}

The model described above predicts the fluctuations in the large scale
optical depth given a set of parameters including $T_{\rm min}$,
$t_{\rm lt}$, $\gamma$, and $R_{\rm mfp}$. By comparing this model to
the data we find constraints on $T_{\rm min}$ using prior information
on the other parameters. To compute the distribution
${dP}/{dT_{\rm min}}$, we first find the likelihood for $T_{\rm
min}$, which may be computed as 
\begin{equation}
L_{T_{\rm min}}\propto\int dR_{\rm mfp}\frac{dP}{dR_{\rm mfp}} \int d\gamma\frac{dP}{d\gamma} \int dt_{\rm lt}
\frac{dP}{dt_{\rm lt}} L_{\delta_\tau}(R_{\rm mfp},\gamma,t_{\rm lt},T_{\rm
min}).
\end{equation}
 Here the a-posteriori distribution for lifetime is computed from
$\frac{dP}{dt_{\rm lt}} \propto \int df_{\rm star} \frac{d^2P}{df_{\rm
star}dt_{\rm lt}}$, and $\frac{dP}{dR_{\rm mfp}}$ is determined from Fan et
al.~(2006). The prior probability distributions for $f_{\rm star}$ and
$t_{\rm lt}$ are taken to be flat in the logarithm. The likelihood
$L_{\delta\tau}(R_{\rm mfp},\gamma,t_{\rm lt},T_{\rm min})$ was computed
from the observed value of the variance in $\delta_\tau$. The uncertainties
on $\delta_\tau$ were computed assuming Gaussian scatter in $\tau_{\rm
eff}$ and a sample size of 19 at each redshift (corresponding to 19
lines-of-sight).  We find the a-posteriori distribution for $T_{\rm min}$
using a flat prior probability distribution in $\log(T_{\rm min})$, i.e.
$\frac{dP}{dT_{\rm min}}\propto L_{T_{\rm min}}\frac{dP_{\rm
prior}}{dT_{\rm min}}$.

\section{Constraints on $T_{\rm min}$ including merger driven star formation}

\begin{figure*}[hptb]
\epsscale{.8}  \plotone{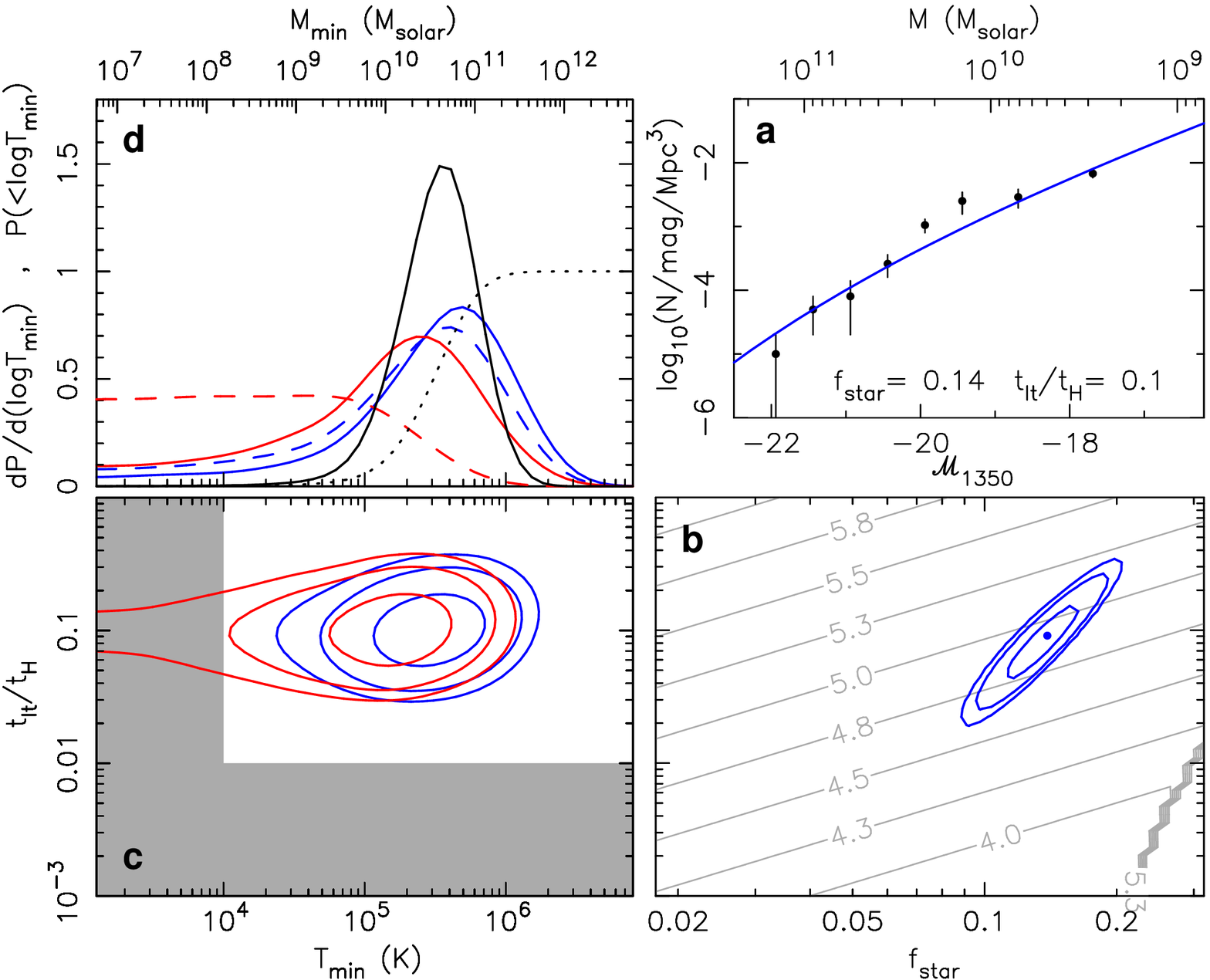}
\caption{\label{fig1s} {\footnotesize Constraints on the star-formation
efficiency ($f_{\rm star}$), duty-cycle ($t_{\rm lt}/t_{\rm H}$) and
minimum virial temperature ($T_{\rm min}$) of galaxies at $z\sim5.5-6$. The
constraints were placed assuming the merger driven model for star
formation. {\bf Panel~a:} The luminosity function inferred from
the UDF, together with the best-fit solution for a luminosity function
model based on the Sheth-Tormen mass function (with luminosity proportional
to halo mass). The upper axis shows the halo mass corresponding to
$\mathcal{M}_{1350}$ for the best fit. {\bf Panel~b:} Contours
(at 64, 26, and 14\% of the maximum likelihood) of the joint a-posteriori
probability distribution $[{d^2P}/{d(\log{f_{\rm star}})d(\log{t_{\rm lt}})}]$. For
comparison, the grey contours show the virial temperature of halos [labeled
using $\log_{10}(T_{\rm vir})$] corresponding the lowest luminosity bin in
the UDF. {\bf Panel~c:} The joint probability distributions for
$\log{(t_{\rm lt}/t_{\rm H})}$ and $\log{(T_{\rm min})}$, which include the constraint on
$t_{\rm lt}/t_{\rm H}$ from the galaxy luminosity function. The contours
represent values at 64, 26 and 14\% of the maximum likelihood, and are
shown for observations at $z\sim5.45$ (blue contours) and $z\sim5.65$ (red
contours). This distribution includes constraints on $t_{\rm lt}$ from the
galaxy luminosity function. However assuming the most likely value of
$R_{\rm mfp}$, we find that $T_{\rm min}\ga10^{5}$K at $z\la5.5$ in all
cases where the duty-cycle of the UV-bright phase of galaxies is greater
than 1\% (corresponding to the lifetime of massive stars). {\bf Panel~d:} Differential probability distributions for $\log{(T_{\rm min})}$ at
$z\sim5.25$ (dashed blue lines), $z\sim5.45$ (blue lines), $z\sim5.65$ (red
lines) and $z\sim5.85$ (dashed red lines). We also show combined
constraints using the lowest three redshift bins (black lines). For
comparison, the upper axis shows the corresponding masses at $z=5.5$. In
this case the solid and dotted lines correspond to differential and
cumulative distributions.  Throughout this work, we adopt the
latest values for the cosmological parameters as inferred from the {\em
Wilkinson Microwave Anisotropy Probe} data.}}
\end{figure*}

For simplicity our fiducial analysis assumed that the star formation rate
is approximately proportional to the derivative of the collapsed fraction
of baryons in halos with a virial temperature $T_{\rm vir}>T_{\rm
min}$. However, starburst episodes are believed to be triggered by galaxy
mergers. We have therefore repeated our calculation of the constraint on
$T_{\rm min}$ with a prescription for star formation that is based on
galaxy mergers rather than the derivative of the collapsed fraction. The
star formation model combines the contribution from mergers of halos with
an initial $T_{\rm vir}<T_{\rm min}$ (that go above $T_{\rm min}$ after the
merger) with the contribution from mergers of galaxies with an initial
$T_{\rm vir}>T_{\rm min}$.  Thus, gas that has not previously cooled and
undergone an episode of star formation in a galaxy forms stars with
efficiency $f_{\rm star}$ as it cools following a merger to form a
galaxy more massive than $M_{\rm min}\equiv M(T_{\rm min})$. In addition,
gas in a galaxy of mass $M>M_{\rm min}$ undergoes an additional starburst
if the galaxy merges with a second galaxy whose mass is larger than
$M/2$. The model assumes one previous burst of star formation in the gas
involved in the major merger so that a fraction $1-f_{\rm star}$ (where $f_{\rm
star}$ is the star formation efficiency) of the gas is available for the
new burst. We assume $f_{\rm star}=0.1$, though the results are quite
insensitive to this parameter. The resulting star formation rate may be
written
\begin{eqnarray}
\nonumber 
\frac{d\Gamma}{dt}(z)&=&\frac{\Omega_{\rm b}}{\Omega_{\rm m}}\int_{M_{\rm
min}/2}^{\infty} dM\int_{\max(0,M_{\rm min}-M)}^{\min(M,M_{\rm
min})}d\Delta M \left.\frac{d^2N_{\rm mrg}}{d\Delta
Mdt}\right|_{M}\frac{dt}{dz}\frac{dn}{dM}f_{\rm star}\left(\Theta(M_{\rm min}-M) M+\Delta M\right)\\
\nonumber &&+\frac{\Omega_{\rm b}}{\Omega_{\rm m}}\int_{M_{\rm min}}^{\infty} dM\int_{M/2}^{M}d\Delta M
\left.\frac{d^2N_{\rm mrg}}{d\Delta
Mdt}\right|_{M}\frac{dt}{dz}\frac{dn}{dM}\\
&&\hspace{10mm}\times f_{\rm star}(1-f_{\rm star})\left[\Theta(M-M_{\rm
min})M + \Theta(\Delta M-M_{\rm min})\Delta M\right], 
\end{eqnarray}
where $\Omega_{\rm m}$ and $\Omega_{\rm b}$ are the fractions of the
critical cosmological density comprised of mass and baryons respectively.
Here $\frac{dn}{dM}(z)$ is the Press-Schechter mass function of dark-matter
halos at redshift $z$, and $\frac{d^2N_{\rm mrg}}{d\Delta
Mdt}\left.\right|_{M}$ is the number of mergers per unit time of halos of
mass $\Delta M$ with halos of mass $M$ (forming new halos of mass
$M_1=M+\Delta M$) at redshift $z$. The function $\Theta(x)$ is the
Heaviside step function. The first term accounts for star formation when
gas first assembles into a halo with $M>M_{\rm min}$. The second term
corresponds to the secondary starbursts (in gas that has already undergone
a burst of star formation) that follow major mergers.

This alternative estimate of the star formation rate leads to results that
are shown in the left panels of Figure~\ref{fig1s} of this supplement. The
preferred value is $T_{\rm min}\sim10^{5.5\pm0.3}$K, with $T_{\rm min}\ga10^{4}$K
at the $\ga99\%$ confidence level.

\section{Constraints on $T_{\rm min}$ including supernovae feedback}

In addition to the suppression of star formation in dwarf galaxies due to a
reionized Universe, there may also be suppression of star formation within
low mass galaxies as a result of supernovae driven-winds which expel gas
from their shallow potential wells. In the local universe, this suppression
is apparent at observed circular velocities $V_{\rm obs}\la178~{\rm
km~s^{-1}}$. A recent study has shown that in a
large sample of local galaxies, the ratio $\epsilon=M_\star/M_{\rm halo}$
(where $M_\star$ and $M_{\rm halo}$ are the total mass of stars and the
dark matter halo respectively) scales as
\begin{eqnarray}
\label{eps}
\nonumber \epsilon&\propto& \left({M_{\rm halo}}/{M_{\rm
halo}^\star}\right)^{2/3}\hspace{5mm}\mbox{ for}\hspace{5mm} M<M_{\rm
halo}^\star\\ &\propto&1\hspace{5mm}\mbox{otherwise}
\end{eqnarray}
The star formation efficiency is proportional to $\epsilon$. 

\begin{figure*}[t]
\epsscale{.8}  \plotone{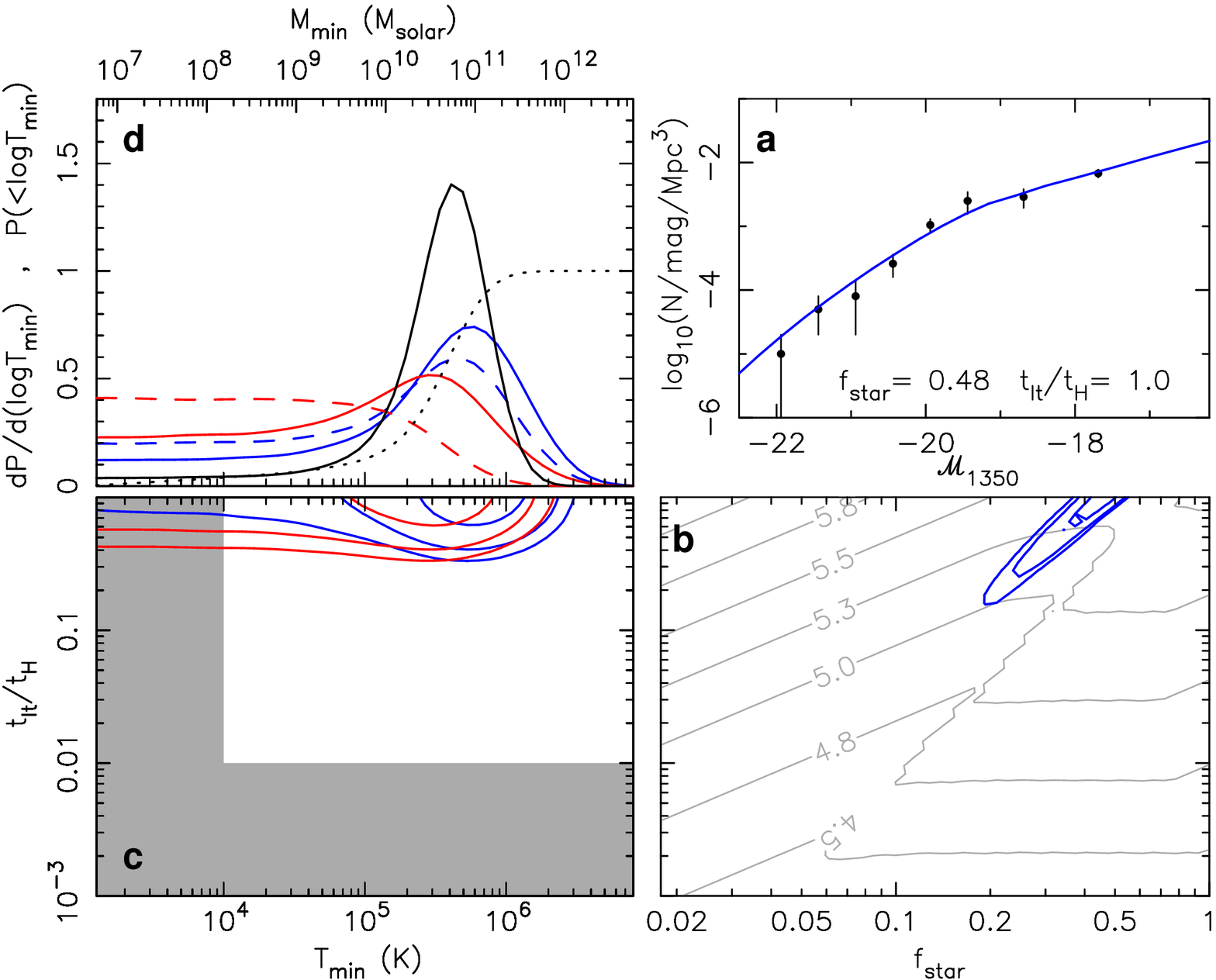}
\caption{\label{fig2s} {\footnotesize Constraints on the star-formation
efficiency ($f_{\rm star}$), duty-cycle ($t_{\rm lt}/t_{\rm H}$) and
minimum virial temperature ($T_{\rm min}$) of galaxies at $z\sim5.5-6$. The
constraints were placed assuming the fiducial model for star formation,
with suppression in dwarf galaxies due to supernovae feedback. The lines
and symbols have the same meaning as in Figure~\ref{fig1s}.}}
\end{figure*}

To explore the effect of supernovae feedback we have repeated our
calculation of the luminosity function with the following modification. We
allow the critical mass $M_{\rm halo}^\star$ at $z\sim5.5$ to be a free
parameter, and fit the luminosity function in equation~(\ref{LF}) of this supplement,
with the derivative $\frac{dM}{d\mathcal{M}_{1350}}$ modified from
the fiducial case because $L\propto\epsilon M_{\rm halo}$. Since
feedback is thought to operate over a fixed fraction of a dynamical
time, we assume the starburst lifetime to be constant among galaxies
of different mass. To avoid discontinuities in the derivative
$\frac{dM}{d\mathcal{M}_{1350}}$, we smooth $\epsilon(M)$ using a
log-normal window function of width 0.25dex.  The best-fit luminosity
function model that includes supernovae feedback is shown in
Figure~\ref{fig2s} of this supplement. We find that allowance for
supernovae feedback suggests high redshift galaxies to be located in
more massive halos, with long duty-cycles $t_{\rm lt}/t_{\rm
H}\ga0.5$.

Estimates of the
scatter in the ionizing background due to Poisson noise in the number of
galaxies and due to fluctuations in the star formation rate resulting from
delayed or enhanced structure formation must also be modified to account
for supernovae feedback if it operates at high redshift. To explore the
effect of supernovae feedback on fluctuations in the ionizing background
and hence on conclusions regarding the suppression of dwarf galaxy
formation in the reionized IGM, we have repeated the comparison of our
modeled fluctuations with the observed fluctuations in $\tau_{\rm eff}$,
including the following modifications. First, in calculating the Poisson
noise we include a galaxy luminosity function that takes account of
supernovae feedback in both the numerator and denominator of
equation~(5). Second, our fiducial model assumes that the derivative of
the dark-matter collapsed fraction multiplied by the star formation
efficiency ($f_{\rm star}$) and baryon fraction, approximates the star
formation rate. We replace the derivative of collapsed fraction in this
formalism with
\begin{equation}
\frac{dF_{\rm col}}{dt} = \frac{d}{dt} \int_{M_{\rm min}}^\infty dM \epsilon M \frac{dn}{dM},
\end{equation}
where $\epsilon$ is computed using equation~(\ref{eps}) of this supplement.
The results are shown in Figure~\ref{fig2s} of this supplement. The most
likely value is $T_{\rm min}\sim10^{5.5^{+0.3}_{-0.4}}$K in agreement with our previous
calculation, while $T_{\rm min}\la10^4$K is ruled out at the $\ga95\%$ level.

\end{document}